\documentclass[journal,onecolumn,12pt]{IEEEtran}

\usepackage{graphicx}

\batchmode

\usepackage{cite}

\usepackage[american]{babel}

\newtheorem{theorem}{Theorem}

\usepackage[T1]{fontenc}% optional T1 font encoding

\ifCLASSINFOpdf
\else
\fi
\usepackage{amsmath}
\usepackage[cmintegrals]{newtxmath}
\begin{document}
%
% paper title
% Titles are generally capitalized except for words such as a, an, and, as,
% at, but, by, for, in, nor, of, on, or, the, to and up, which are usually
% not capitalized unless they are the first or last word of the title.
% Linebreaks \\ can be used within to get better formatting as desired.
% Do not put math or special symbols in the title.

\title{Optimization of Virtual Networks}

%
%
% author names and IEEE memberships
% note positions of commas and nonbreaking spaces ( ~ ) LaTeX will not break
% a structure at a ~ so this keeps an author's name from being broken across
% two lines.
% use \thanks{} to gain access to the first footnote area
% a separate \thanks must be used for each paragraph as LaTeX2e's \thanks
% was not built to handle multiple paragraphs
%

%\author{Andr\'as Farag\'o,~\IEEEmembership{Senior Member,~IEEE}
        % <-this % stops a space
%\thanks{A. Farag\'o is with the Department
%of Computer Science, The University of Texas at Dallas, TX 75080, USA,
%e-mail: farago@utdallas.edu}% <-this % stops a space
%}

\author{Andr\'as Farag\'o

\thanks{This work has been submitted to the IEEE for possible publication. Copyright may be transferred without notice, after which this version may no longer be accessible.}
Department of Computer Science \\ The University of Texas at Dallas \\ 
Richardson, TX 75080, USA \\ {\tt farago@utdallas.edu}
}

\maketitle

% As a general rule, do not put math, special symbols or citations
% in the abstract or keywords.
\begin{abstract}
We introduce a general and comprehensive model for the design and optimization of  Virtual Networks, and the related concept of Network Slicing.
The model is flexible, so that by adjusting some of its elements, it can accommodate  many different specific cases
of importance. Yet, surprisingly, it still allows efficient optimization, in the sense that the global optimum can be well approximated
by efficient algorithms.
\end{abstract}

% Note that keywords are not normally used for peerreview papers.
%\begin{IEEEkeywords}
%Communications Society, IEEE, IEEEtran, journal, \LaTeX, paper, template.
%\end{IEEEkeywords}

% For peer review papers, you can put extra information on the cover
% page as needed:
% \ifCLASSOPTIONpeerreview
% \begin{center} \bfseries EDICS Category: 3-BBND \end{center}
% \fi
%
% For peerreview papers, this IEEEtran command inserts a page break and
% creates the second title. It will be ignored for other modes.
\IEEEpeerreviewmaketitle

\section{Introduction}
% The very first letter is a 2 line initial drop letter followed
% by the rest of the first word in caps.
% 
% form to use if the first word consists of a single letter:
% \IEEEPARstart{A}{demo} file is ....
% 
% form to use if you need the single drop letter followed by
% normal text (unknown if ever used by the IEEE):
% \IEEEPARstart{A}{}demo file is ....
% 
% Some journals put the first two words in caps:
% \IEEEPARstart{T}{his demo} file is ....
% 
% Here we have the typical use of a "T" for an initial drop letter
% and "HIS" in caps to complete the first word.
\IEEEPARstart{T}{he} general concept of {\em virtualization} refers to the creation of a logical structure that is distinct 
from the underlying physical infrastructure. 
In the networking field virtualization plays an important role: virtual networks already have a long history. One of the first precursors
 was when companies could 
connect various sites through leased lines over the public switched telephone network, thus forming an early, still primitive, embodiment of a 
{\em Virtual Private Network (VPN).} The early data networks, such as X.25 in the 1970s, Frame Relay in the 1980s, and Asynchronous Transfer Mode
(ATM) in the 1980-90s,  already supported VPN-like configurations. {\em Virtual Local Area Networks (VLANs)} also emerged in the 1980s; the IEEE later standardized VLANs over Ethernet in the IEEE 802.1Q standard. In the Internet, some efforts were present from the very beginning to allow 
the logical separation of some part of the traffic, or to provide dedicated treatment to a traffic class. This intent was first represented by the 
8-bit Type of Service field of the IPv4 header, which later became Differentiated Services Code Point (DSCP), to support Differentiated Services.
These could also be viewed as primordial early precursors of Network Slicing (see below). Virtual Networks could  be configured via 
Multiprotocol Label Switching (MPLS), as well, which came into view in the 1990s. 
In the modern Internet, initially the possibility of flexible experimentation with new protocols and architectures motivated the concept of {\em Overlay Networks}, which are a type of Virtual Networks, see Anderson, Peterson, Shenker, and  Turner \cite{anderson}. By now,  the Virtual Network concept became ubiquitous. 

A related, but not fully identical, concept is {\em Network Slicing}, which emerged in connection with fifth generation (5G) wireless networks,
and created quite a bit of excitement in the cellular network research community. This is based on perceiving the concept as a  revolutionary new 
way to extending network capabilities, in combination with the  evolutionary improvement of efficiency. To illustrate it, 
let us quote from the survey article of Foukas et al.\ \cite{foukas}:

\begin{quote}
``What 5G systems are going to be has yet to
be determined. However, it is conceivable that
the eventual 5G system will be a convergence
of two complementary views that are currently
driving the research and industrial activity on 5G.
One is an {\bf\em evolutionary} view focusing on significantly
scaling up and improving the efficiency
of mobile networks (e.g., 1000x traffic volume,
100x devices, and 100x throughput). Much of
the research focused around this view is on the
radio access side looking at novel technologies
and spectrum bands (e.g., massive multiple-input
multiple-output, MIMO; millimeter-wave).
The other {\bf\em service-oriented} view envisions 5G
systems catering to a wide range of services differing
in their requirements and types of devices,
and going beyond the traditional human-type
communications to include various types of
machine-type communications. This requires the 
network to take different forms depending on the
service in question, leading naturally to the notion
of {\bf\em slicing} the network on a per-service basis (...) Realizing this service-oriented
view requires a radical rethink of the mobile network
architecture to turn it into a more flexible
and programmable fabric, leveraging technologies
like software defined networking (SDN) and
network functions virtualization (NFV), which can
be used to simultaneously provide a multitude of
diverse services over a common underlying physical
infrastructure." 
\end{quote}

\noindent

It is natural to ask at this point: 
{\em how do Network Slices and Virtual Networks differ?} There is clearly significant overlap between the two concepts, since they both consist of a 
{\em logically} separated subset of network resources, dedicated to serve some demand. On the other hand, in case of Virtual Networks  
this aims mostly at serving a customer, such as setting up a VPN for a company. Therefore, it is typically {\em horizontal} logical separation.
In a Network Slice, the key goal is to logically separate resources for {\em services,} at different levels, including cloud, fog, edge, 
processing power and storage capacity, switches, base stations, radio
access network etc, so that they together can best provide the resources for the service. Thus, in this sense, the Network Slice has a more  
{\em vertical} focus. 
%Network Slicing is illustrated in the figure below with three slices.

%\includegraphics[scale=0.45]{fig0.pdf}

It is not difficult to believe that the need for designing, configuring, dimensioning, as well as physically embedding  Network Slices and Virtual Networks leads to  hard problems. We review some issues in the next section.

%\section{Intellectual Merit}

\subsection{Previous Work}

The most investigated task in connection with Virtual Networks is how to {\em embed} them into a {\em substrate network.} The substrate is 
most often an actual physical network, but it may also be another virtual network. This embedding entails a number of 
sub-tasks, that may be logically separated, or can also be handled in a more complex {\em joint optimization} model. The most important sub-tasks are: 
\begin{itemize}\itemsep=1mm

\item 
Assigning the virtual nodes into physical ones, while taking into account processing demands and physical node capabilities, as well as various preferences, such as geographic distance. 

\item 
Setting up physical routes to implement the logical links. This may encompass disjointness constraints, and various cost and reliability considerations. 

\item Dimensioning the logical links, i.e., assigning transmission capacities to them along the physical routes, taking traffic related issues, 
such as  blocking probabilities into account.

\end{itemize}
These tasks have been intensely researched for more than a decade. Many variants have been investigated. For example, 
the underlying physical network may be static, but may also be dynamically reconfigurable. The latter happens, e.g., in data center networks,
see Curran et al.\ \cite{curran}. The demand for Virtual Networks, and their parameters, may also be static, or stochastic. 
 The 2013 survey paper by Fischer et al.\ \cite{fischer} already reviews as many as
80 published algorithms, and since the time the survey has been published, many more have been proposed. (More than 1,100 papers refer to this 
survey!) Some of the
approaches solve the various sub-tasks separately, while others set up a larger joint optimization problem, and search for a global optimum. 
However, a common feature in {\em all cases} that one has to solve {\bf NP}-hard optimization tasks, due to the involvement of {\bf NP}-complete
problems, which are known to be notoriously hard. As a result, the proposed solutions either lead to prohibitively long computation time, 
or are forced to give up  optimality. 

The Network Slicing literature is somewhat smaller, since it is a more recent concept. There are attempts to formalize it in various ways: as a complex 
optimization problem (Han et al.\ \cite{han}), to approach it with the tools of machine learning (Le et al.\ \cite{le}), using game theory 
based auction models (Jiang et al.\ \cite{jiang}), big data analytics (Raza et al.\ \cite{raza}), and a number of others. Again, a common feature is that hardness unavoidably forces the algorithms to sacrifice either speed or optimality. Furthermore, there is not even clear agreement on what is the ``right" 
model that can best capture the problem. 
With some sarcastic exaggeration, we could describe the status of research with this statement: 
we do not know how to solve the problem, but we are not even sure what exactly to solve.  

Thus, by and large, we can characterize the current situation in Virtual Networks and Network Slicing with the following features:
\begin{itemize}\itemsep=0mm

\item There is no clear consensus  on how to capture the problems, which model is best for which situation.

\item The algorithmic solutions are either too slow, or else they sacrifice  optimality. In the latter case the authors  resort to various heuristics, 
typically without any {\em performance guarantee.} Nevertheless, this is usually accepted with the argument ``what else can you do?"

\end{itemize}

\subsection{Our Goal}

The above outlined situation motivates us to propose a new approach, with the following main features:

\begin{itemize}\itemsep=1.5mm

\item We intend to make our model general and comprehensive enough, so that it can capture many special cases. 
%both in Network Slicing and in Virtual Network configuration.  
In this sense, we do not want to propose 
{\em just another} model, after hundreds of previous ones. Rather, we aim at creating a common generalization of many special cases, applicable both to Virtual Networks and Network Slicing. The specific problem we focus on is this: how much capacity should be allocated to each virtual/logical entity, such that 
the total carried traffic (or its weighted version, the network revenue) is maximized, under the constraint that the logical capacities together fit in the available physical capacities. 

\item We intend to make the model flexible, so that by adjusting some of its elements, it can accommodate  many different specific cases
of importance.

\item The hardest part is this: what to do with the evergreen desire for optimality? We cannot reasonably aim at optimally solving {\bf NP}-hard optimization problems 
efficiently. On the other hand, we are also not satisfied by resorting 
to mere heuristics without any performance guarantee, that is, finding ad-hoc solutions that can be arbitrarily far from the optimum. But is there anything
else that one can reasonably do? We believe the answer is yes, but in a quite nontrivial way. We elaborate on it in the subsequent sections.

\end{itemize}

\section{The Proposed Model}
\label{sec2}

\subsection{Elements of the Model}
\label{elements}

\begin{itemize}\itemsep2mm

\item {\bf Physical Entities.} They represent the {\em physical} parts of the network, such as nodes, links, computers, routers, storage units, base stations, sensors, etc.,  i.e, {\em anything} that is physically present and takes part in the operation of the network. We denote the physical entities by $P_1, \ldots,P_n$.
%and collectively they form the set ${\cal P}=\{P_1,\ldots,P_n\}$.

\item {\bf Physical Capacities.} Each physical entity is represented by a numerical parameter, called its {\em capacity.} The unit in which the capacity is measured depends on the nature of the physical entity. For example, if it is a link, then its capacity means its maximum transmission rate, which can be measured, say, in Mb/s. If it is a processor, then its capacity is the processing speed in some unit, such as megaflops. If it is a storage device, then its capacity is its storage capacity, expressed, say, in gigabytes. If it is a radio link,
then its capacity is bandwidth, which can be measured in MHz.

\item {\bf Capacity Types.}
The physical nature of the capacity parameter (whether it is transmission rate, bandwidth, processing speed, storage size, etc) is referred to as the {\bf type} of the capacity. 
%We denote the types by $T_1,\ldots,T_k$, where $k$ is the number of different types in the network. 
%The type of physical entity $P_i$ is denoted by $T(P_i)$. We also use a vector that collects these types:
%$${\bf T}({\cal P})=(T(P_1), \ldots,T(P_n)).$$
 
\item {\bf Vector of Physical Capacities.} 	We collect the physical capacity values in a vector, denoted by ${\bf C}_{phys}$. That is, the 
$i^{th}$ coordinate of  ${\bf C}_{phys}$ is the capacity of physical entity $P_i$, $i=1,\ldots,n$.

\item{\bf Physical Entities with Multiple Parameters.} A network can naturally contain entities that cannot be represented by a single parameter.
For example, a computer can have a certain processing speed and a certain storage capacity. To capture this, we could certainly represent the capacity
by a vector. That, however, would make the notation messy, as these vectors could have a different number of components, but we still want 
to collect them in ${\bf C}_{phys}$. Therefore, 
we rather reduce it to the single parameter case,
by representing the physical entity as several entities, one for each parameter. 
In this example, the computer would be represented as two physical entities: a processor with some processing speed, and a storage unit with some storage capacity.

\item {\bf Logical Entities.} Any {\em subset} of physical entities with the same capacity type can form a {\em logical entity.} For example, if a logical link $\cal L$ is embedded in a network by a physical route, containing the physical links $L_1,\ldots,L_k$, each characterized by its transmission 
rate, then they together form a logical entity. Note that different logical entities, as subsets, are allowed to overlap. 

It is also allowed that the set contains only a single physical entity. For example, a logical node, such as a Virtual Machine, may be mapped into a single physical node. Such Virtual Machines can carry out various service functions for packets, such as firewall, proxy server, they can perform deep packet inspection, access control, network address translation, traffic compression, QoS policy enforcement, traffic optimization, etc. This allows implementing the 
functionality of {\bf\em Service Function Chaining.} We denote the logical entities by ${\cal L}_1,\ldots,{\cal L}_m$.  

\item {\bf Incidence of Logical and Physical Entities.} This is represented by an  $m\times n$ matrix ${\bf S}=[s_{ij}]$  with 0-1 entries. 
It expresses the incidence of logical and physical entities such that the $i^{th}$ row indicates which physical entities are contained 
in (i.e., used by) logical entity ${\cal L}_i$. That is, 
\begin{eqnarray} \nonumber
s_{ij} = \left\{
    \begin{array}{ccl}
           1  & \;\;\mbox{ if }\;\;\; P_j\in {\cal L}_i \\
           0  & \;\;\; \mbox{ if }\;\;\; P_j\notin {\cal L}_i.
    \end{array} \right.
\end{eqnarray}

\item {\bf Logical Capacities.} Each logical entity, just like the physical ones, is characterized by a parameter called {\em logical capacity}.
This has the same type as the physical capacities of the physical entities that make up the logical entity (these physical entities  are required to have the same type of capacity).
%The type of logical entity ${\cal L}_i$ is denoted by $T({\cal L}_i)$.

\item {\bf Vector of Logical Capacities.} Let us denote the capacity of logical entity ${\cal L}_i$ by $C_i$. We collect them in a vector,
denoted by ${\bf C}=(C_1,\ldots,C_m)$.

\item {\bf Capacity Constraints.} The summed capacities of the logical entities that use a given physical entity cannot be more than 
the capacity of the physical entity (note that they all must have the same type, due to the shared physical entity). This holds for every physical entity, which can be concisely expressed in matrix-vector notation as 
\begin{equation}\label{SC}
{\bf S}{\bf C}\leq {\bf C}_{phys}.
\end{equation}

\item {\bf Offered Load to a Logical Entity.} The offered load represents the {\em demand} for a logical entity. For example, if it is a logical link, then the demand is the offered traffic. For processors it may 
be the requested processing speed. For storage units, the offered load is the data amount to be stored.  The demand can be random, represented by a stochastic process, which is typically the case, e.g., for the offered traffic load.
We assume, for (initial) simplicity, that this stochastic process has a time-invariant expected value, denoted by $\rho_i$, which is the offered load to logical entity ${\cal L}_i$.

\item{\bf Loss Functions.} If a logical entity ${\cal L}_i$ receives an offered load of $\rho_i$, then it may not be able to fully satisfy this 
demand. We say that {\em loss} occurs, in the sense that the {\em carried load}, the part of the demand that is actually satisfied, is less
than the offered load. The carried load is also captured (for initial simplicity) by a time-invariant expected value.
If $\rho_i>C_i$, i.e., the demand exceeds the capacity, then loss must clearly occur. But even if $\rho_i\leq C_i$, loss can still occur, due to the 
possible random fluctuation of the offered load around its average. 

We capture the loss by a {\em loss function}, denoted by  $F_i(\rho_i, C_i)$.
The meaning of the loss function is that if logical entity ${\cal L}_i$ receives an offered load $\rho_i$, and has capacity $C_i$, then  
an $F_i(\rho_i, C_i)$ fraction of the offered load is lost. We can also view it as loss probability or blocking probability. As a result, the carried load will be
$\rho_i (1-F_i(\rho_i, C_i))$.

These loss functions may be different for different logical entities, that is why they are indexed by $i$. Note that even though the arrival process 
(offered load) is characterized by its expected value $\rho_i$, the choice of the loss function can express much more information about
the detailed  stochastic behavior. After all, $F_i(\rho_i, C_i)$ can represent the probability that the {\em actual} random load exceeds a threshold $C_i$. In this sense,
the values of $F_i(\rho_i, C_i)$ with different $\rho_i, C_i$ values can describe the entire probability distribution of the load (assuming it is stationary, 
which we assume for initial simplicity).
To provide maximum model flexibility, we allow  arbitrary real-valued functions for the $F_i$, except that they are required to satisfy the following 
basic conditions, which are needed for making the mathematical analysis feasible:

\vspace*{-0.2mm}
\begin{description}\itemsep0.5mm

\item[\rm (i)\,\,] 
$0\leq F_i(\rho_i, C_i) \leq 1$, so that we can view it as loss probability.

\item[\rm (ii)\,\,] 
$F_i(\rho_i, C_i)$ is a continuous function in both variables. 

{\em Note:} the stronger property of being differentiable is not required. For example, the function may have breakpoints where the 
derivative does not exist. 

\item[\rm (iii)\,\,] 
$F_i(\rho_i, C_i)$ is increasing in $\rho_i$, for  any fixed $C_i$, and decreasing in $C_i$, for  any fixed $\rho_i$.
(the increase/decrease does not have to be strict, the function may remain constant). In words, these express the natural expectation
that putting more load on the same capacity cannot result in smaller loss, and 
adding more capacity to carry the same load cannot increase the loss. 

\end{description}

\item {\bf Flows.} The definition of a flow has two parts: (1) a set of logical entities; we say that the flow {\em traverses} these entities;
and (2) it is characterized by the following parameters: 
\begin{itemize}

\item {\bf Offered flow:} A flow amount that we would like to push through.

\item {\bf Capacity demand:} A capacity demand per unit offered flow for each traversed logical entity. 

\end{itemize}

To understand the meaning of flow amount, let us  bring a classical analogy. Imagine that we want to use a route for broadband calls in the 
telephone network. Then the offered flow is the number of such calls we would like carry by the route. The capacity demand tells how many 
circuits are needed on each link for one call.

We generalize this classical scenario in several ways. First, the flow can be served by an arbitrary set of entities, not only by a route. 
Second, these entities
can have different types of capacities. For example, transmission rates of links, processing capacities of nodes, storage capacity of storage units, 
bandwidth of radio links, etc.
Three, the amount of capacity needed for one unit of flow may be different on different entities. Four, these are logical entities. For example,
logical links that may be implemented by physical routes.
(Keep also in mind, however, that a logical link/node may contain a single physical entity; so we can also represent the case when the components are physical.) 

\item {\bf Virtual Networks or Network Slices.} A Virtual Network or Network Slice is defined as a system of flows. (For initial simplicity, 
we first consider a static set.) Observe that we have a nested 
system of abstractions here. For example, a Network Slice is described by a set of flows, each flow incorporates a subset of logical entities, and each 
logical entity is made up by a subset of physical entities. Of course, multiple network slices may exist in the network simultaneously, and share the
underlying physical capacities. 

\item {\bf Goal.} The key goal of the optimization is to tell how much capacity should be allocated to each logical entity, such that 
the total carried traffic (or a weighted version of thereof) is maximized, under the constraint that the logical capacities together fit in the available physical capacities. 

{\bf Important note:} the loss functions can reflect very different requirements. Allowing this is a key feature of the model. For example, one loss function
 can express classical blocking probability. Another one can express, e.g., that a service, such as emergency notification, needs resources which provide extremely low delay and high reliability, or else its loss will be unacceptably high. Yet another service, such as video conferencing, may need resources that guarantee low jitter, small packet loss, and high bandwidth, or else its quality will be unacceptable, leading to high loss. 
 It is a main feature of the model that we can treat 
 these very different requirements in a unified way, without losing the ability of efficient optimization.

\item {\bf Notational convention:} To make the notation easier to follow, whenever it does not cause confusion, we denote an entity simply by its index. 
In this vein, let us number the flows that exist in the network by $1,\ldots,R,$ 
and the logical entities by $1,\ldots,m$. For every $j,r$, let $A_{jr}$ denote the demand (capacity units) that is requested on
logical entity $j$ by flow $r$, per unit offered flow. We assume that $A_{jr}$ is integer valued, which can always be achieved by an appropriate 
scaling of units. If the flow does not traverse $j$, then $A_{jr}=0$. Further, let us denote the offered load of flow $r$ by $\nu_r$.

\end{itemize} 

\subsection{Fundamental Equations}
\label{fundeq}

For (initial) simplicity we introduce the assumptions below, because they significantly help the analysis.
 These assumptions lead to asymptotically exact results, i.e., 
only cause vanishing errors, when the capacities grow large, as was analyzed under classical scenarios, see Kelly \cite{kelly,kelly2}, Labourdette \cite{labour}. 

\smallskip
\noindent
{\bf Independence Assumptions:} 
\begin{enumerate}\itemsep1mm

\item 
The losses on different logical entities are considered 
independent random events.  

\item When multiple units of capacity are needed for a unit of flow on a logical entity, it is modeled 
as  grabbing each capacity unit independently, if available.

\end{enumerate}

\smallskip

Let us now compute the carried load of flow $r$. It has an offered load $\nu_r$. On each logical entity $j$ that the flow traverses
it suffers a (relative) loss of $F_j(\rho_j,C_j)$, where $F_j,\rho_j,C_j$ are the loss function, offered load, and capacity of logical entity $j$, 
respectively. That is, the probability that the flow can successfully grab a unit of available capacity on $j$ is 
$1-F_j(\rho_j,C_j)$. Since each unit of  flow needs $A_{jr}$ units of capacity, therefore, by the second independence assumption, 
its success probability on $j$ (the probability that the unit of flow gets through) will be $\big( 1-F_j(\rho_j,C_j)\big)^{A_{jr}}$. Then we can obtain the success
probability of a unit of the entire flow, as the product of these probabilities (by the first independence assumption) over the set of all logical entities 
${\cal F}_r$ that flow $r$ uses:
$$\prod_{j\in {\cal F}_r} \Big( 1-F_j(\rho_j,C_j)\Big)^{A_{jr}}.$$
Observe now that whenever $j\notin {\cal F}_r$, , we have $A_{jr}=0$, leading to $\big( 1-F_j(\rho_j,C_j)\big)^{A_{jr}}=1$. These factors of 
1 do not change the product value, so we can take the product over all $j$, rather than just $j\in {\cal F}_r$. Multiplying it with the offered load of flow $r$ we get the {\em carried load} of flow $r$ as
\begin{equation}\label{carried}
\nu_r\prod_j \Big( 1-F_j(\rho_j,C_j)\Big)^{A_{jr}}.
\end{equation}

Consider now a logical entity $i$. Recall that if its offered load is $\rho_i$, then its carried load will be $\rho_i (1-F_i(\rho_i, C_i))$,
where $F_i$ is its loss function, and $C_i$ is its capacity. But all the carried load on $i$ must come from the flows that use $i$. Therefore, 
if we sum up the carried loads of all flows, taking into account that flow $r$ uses $A_{ir}$ capacity on $i$ per unit load, then 
we get the following equation:
$$\rho_i (1-F_i(\rho_i, C_i)) = 
\sum_{r} A_{ir} \nu_r\prod_j \Big( 1-F_j(\rho_j,C_j)\Big)^{A_{jr}}.$$
Note that whenever $r$ does not use $i$, we have $A_{ir}=0$, so it is safe to do the summation for all flows $r$.
The above equation holds for every logical entity $i$. Therefore, after rearranging, we get the following system of 
equations\footnote{This system is a generalization of what is known as Erlang Fixed Point Approximation, see Kelly \cite{kelly3}.}
for the offered loads $\rho_i,\, i=1,\ldots,m,$ of the $m$ logical entity:
$$\rho_i = (1-F_i(\rho_i, C_i))^{-1} 
\sum_{r} A_{ir} \nu_r\prod_j \Big( 1-F_j(\rho_j,C_j)\Big)^{A_{jr}}.$$
If the logical capacities $C_1,\ldots,C_m$ are known, then the $\rho_i$ values can be computed from this system by iterated substitution.
Once $\rho_1,\ldots,\rho_m$ are computed, we get the total carried load, by summing up (\ref{carried}) for all flows. Let us denote the 
total carried load by $T$, then we have
$$ T= \sum_r \bigg(\nu_r\prod_j \Big( 1-F_j(\rho_j,C_j)\Big)^{A_{jr}}\bigg).$$
However, the logical capacities $C_j$ are not known! They are precisely what we want to optimize, so that we can tell what the optimal allocation is
of logical capacities, {\em within the physical constraints.} The latter are expressed by the linear system of equations (\ref{SC}).
Note that adding more logical capacity to a logical entity can only happen at the expense of others, since the physical limits are given.

Thus, we face the following, rather complicated looking,  optimization problem:

\begin{equation}\label{opt}
\mbox{Maximize}\;\;\;\; T({\bf C})= \sum_r \nu_r\prod_j \Big( 1-F_j(\rho_j,C_j)\Big)^{A_{jr}}
\end{equation}
\hspace*{6mm}Subject to
\begin{equation} \label{rhoi}
\rho_i = (1-F_i(\rho_i, C_i))^{-1} 
\sum_{r} A_{ir} \nu_r\prod_j \Big( 1-F_j(\rho_j,C_j)\Big)^{A_{jr}}, 
\end{equation}
$$\hspace{50mm} i=1,\ldots,m$$
\begin{equation} \label{lin1}
\;\;\;\; {\bf S}{\bf C}\leq {\bf C}_{phys}
\end{equation}
\begin{equation} \label{lin2}
{\bf C}\geq 0
\end{equation}

\section{Optimization}
\label{optim}

Clearly, the optimization task described above  by (\ref{opt}), (\ref{rhoi}), (\ref{lin1}), and (\ref{lin2}), is quite complex and heavily nonlinear. At first, it appears hopeless to find (or at least closely approximate) a  {\em globally optimal} solution. Surprisingly, however, the objective function  can be well approximated by a {\em concave} function, which is much easier to maximize. Specifically, we can prove a theorem presented below.
To explain it, let us first introduce a useful concept:

\medskip
\noindent 
{\bf Asymptotically concave function.}
A function $f$ is called {\em
asymptotically concave} if for each
$x\geq 1$ the point-wise limit
$$\tilde{f}(x)=\lim_{n\rightarrow\infty} \frac{1}{n}f(nx)$$
exists and $\tilde{f}$ is a concave function in the ordinary sense.

\smallskip

In other words, an asymptotically concave function is {\em almost concave} for large variable values, which is the case we consider (large capacities).

\medskip

\begin{theorem}\label{thm1}
{\em 
If the independence assumptions\footnote{See at the beginning of Section~\ref{fundeq}.} hold, then there exist a correction function $\epsilon({\bf C})$ with the following properties:
\smallskip
\begin{description}\itemsep1.5mm

\item[\,\,\,\,\,(i)] 
$\widetilde T({\bf C})= T({\bf C})+\epsilon({\bf C})$ is an asymptotically concave function of $\bf C$.

\item[\,\,\,\,\,(ii)] 
The correction function
$\epsilon({\bf C})$ is small in the following sense:
$0\leq \epsilon({\bf C})\leq \sum_i \rho_iB_i$, where $B_i=F(\rho_i,C_i)$ is the blocking probability (loss) of logical entity $i$.

\end{description}
}
\end{theorem}
\medskip

{\em Note:}  Observe that $\rho_iB_i$ is the blocked load on logical entity $i$. Since under normal operation the overall blocked load  is expected to be small, therefore, we can expect $\epsilon({\bf C}) \ll T({\bf C})$, resulting in a small difference between the original objective function $T({\bf C})$
and its modified version $\widetilde T({\bf C})$. Also note that all this is valid for {\em any} system of loss functions, as long as they satisfy the
mild mathematical requirements outlined in Section~\ref{elements}.

%%%%%%%%%%%%%%%%%%%%%%%%%%

\medskip
\noindent {\bf Proof of Theorem 1:} See in Appendix A.

\medskip
%%%%%%%%%%%%%%%%%%%%%%%%%%

Theorem~\ref{thm1} gives hope to find an approximation of the  global optimum, since globally maximizing a concave function over a convex domain is a well solved
problem. In our case, the modified objective function $\widetilde T({\bf C})$ is indeed concave. Regarding the constraints, the linear 
inequalities  (\ref{lin1}),  (\ref{lin2}) alone would indeed define a convex domain. Unfortunately, however, 
this is badly messed up by the heavily nonlinear system (\ref{rhoi}) of equations. Nevertheless, surprisingly again, we can prove the following:
%(Farag\'o \cite{tech}):

\medskip
\begin{theorem}\label{thm2}
{\em 
If the independence assumptions hold, then there exist a function $\phi({\bf C})$ with the following properties:
\smallskip
\begin{description}\itemsep1.5mm

\item[\,\,\,\,\,(i)] 
$\phi({\bf C})$ is an asymptotically concave function of $\bf C$.

\item[\,\,\,\,\,(ii)] 
The value of $\phi({\bf C})$ can be computed by a polynomial time algorithm.

\item[\,\,\,\,\,(iii)] 
If ${\bf C}^*$ is an optimal solution of the new optimization problem 
\begin{equation}\label{phi}
\mbox{\rm maximize} \;\;\;\; \phi({\bf C})
\end{equation}
\hspace*{30mm} {\rm subject to} \hspace*{2mm} $${\bf S}{\bf C}\leq {\bf C}_{phys},\;\;\; {\bf C}\geq 0$$

\smallskip
then ${\bf C}^*$ is also an optimal solution to the modified version of the {\em original} optimization problem
$$\mbox{\rm maximize} \;\;\;\;\widetilde T({\bf C}) $$
\hspace*{60mm} {\rm subject to} \hspace*{4mm} {\rm (\ref{rhoi}), (\ref{lin1}),  (\ref{lin2}).}
\end{description}
}
\end{theorem}

\medskip
\noindent {\bf Proof:} See in Appendix A.

\medskip

Thus, we may say that the new objective function $\phi({\bf C})$ can ``swallow" the badly nonlinear system (\ref{rhoi}) of constraints,
and leaves only the linear part to be considered. Yet (and this is the surprising part!) $\phi({\bf C})$ still can be chosen such that it remains 
an efficiently computable asymptotically {\em concave} function.
Once we have it, we see that for large capacities the new optimization task (\ref{phi}) requires only the maximization of an (efficiently computable) 
nearly concave function 
over a convex domain, given by linear inequalities. This optimization task can already be solved globally and efficiently by standard methods of convex optimization, for which off-the-shelf commercial software is also available.

\subsection{Interlude: A Bold Conjecture About Optimization  }
\label{conj}

The above theorems suggest that even a very complicated-looking optimization problem may be approximated by one that is efficiently (i.e., polynomial-time) solvable. (Convex optimization is known to be solvable in polynomial time, see, e.g.,  \cite{bertsekas,bubeck}.) One may wonder: is it only good luck in the considered case that complicated optimization can be approximated efficiently, or is 
it perhaps the manifestation of a more general phenomenon?

Our recent papers \cite{AAIM,Theo} strongly suggest that it is indeed a more general phenomenon. We do not have space here to elaborate the rather complex details (they are detailed
in the referenced papers), so let us just briefly state the essence. We were able to formally prove that for a large 
class of decision problems (i.e, questions with  a yes/no answer) the following holds: every problem in the class can be approximated by a polynomial-time solvable one, in the sense 
that they differ only on an asymptotically vanishing subset of instances. One might view it as an analogy  to the classical Weierstrass Theorem in real analysis:

\begin{quote}
{\em Every continuous function on a bounded interval can be arbitrarily well approximated by a polynomial. In other words, the polynomials constitute a 
{\em dense subset} of all continuous functions over the interval. }
\end{quote}

Our result can be formulated analogously this way: with an appropriate,
more sophisticated  definition of density, the polynomial-time solvable decision problems are dense in a much larger  class of decision problems. What is very surprising,  all known {\em natural} {\bf NP}-complete problems are in the considered class! {\em Natural} means here that it has been studied on its own right, rather than having been constructed artificially (e.g., by diagonalization), just for the sake of an example or counterexample. 

The above result of ours provides strong motivation to conjecture that this phenomenon may carry over to optimization tasks from decision problems. 
Let us informally state the conjecture:

\begin{quote}
\noindent 
{\bf Conjecture 1:} {\em There is a (reasonable) definition of density, such that in the set of natural optimization problems the ones that are
solvable in polynomial time constitute a dense subset. In other words, all natural optimization tasks can be well approximated by efficiently solvable ones.}  
\end{quote}

\smallskip

While the above conjecture may sound very bold and surprising, our results in \cite{AAIM,Theo}
 suggest that it has a quite reasonable chance to hold.
Note that if the conjecture is indeed true, and one can find a constructive proof, then it would have a {\em huge} impact.
It would mean that all the notoriously hard natural optimization problems can be well approximated with efficient algorithms! But this takes quite a bit
of more work, we do not intend to engage into it in the present paper.

\section{Treating Reconfigurable Physical Networks}
\label{reconf}

In some cases the physical network is not fixed, it is reconfigurable. An example is 
when in a data center network the racks of computers communicate via free space laser links, which can be 
quickly reconfigured, when needed, see Curran et al.\ \cite{curran}. Interestingly, this case can also fit in our model.

Consider the case, when the $n$ physical entities of our model are not fixed in advance. Rather, they can be chosen arbitrarily from a set of $N$ {\em potential} physical entities. Let us look at the simplest case, when all potential physical entities have unit capacity, measured in relative units. The logical entities, and the flows on top of them,  can now use all the potential physical entities. Of course, eventually only 
those among them can actually operate, which only use the chosen physical entities, i.e., the ones that are selected from the potential set to come into existence. Can we still fit this in the optimization? The answer is yes! Let ${\bf e}_N$ denote the $N$-dimensional vector in which all components are 1.
We treat now the physical capacity vector ${\bf C}_{phys}$ as a {\em variable,} not constant. Its value  may be any vector in which 
each coordinate is in the interval $[0,1]$, expressing the values in relative units, with a maximum of 1. Then we can modify our optimization task to

\begin{equation}\label{opt'}
\mbox{Maximize}\;\;\;\; T({\bf C})= \sum_r \nu_r\prod_j \Big( 1-F_j(\rho_j,C_j)\Big)^{A_{jr}}
\end{equation}
\hspace*{20mm}Subject to
\begin{equation} \label{rhoi'}
\;\;\; 
\rho_i = (1-F_i(\rho_i, C_i))^{-1} 
\sum_{r} A_{ir} \nu_r\prod_j \Big( 1-F_j(\rho_j,C_j)\Big)^{A_{jr}},\;\;\;i=1,\ldots,m
\end{equation}
\begin{equation} \label{lin1'}
\;\;\;\; {\bf S}{\bf C}\leq {\bf C}_{phys}
\end{equation}
\begin{equation} \label{lin2'}
{\bf C}\geq 0, {\bf C}_{phys}\geq 0
\end{equation}
\begin{equation} \label{lin3'}
{\bf e}_N\geq {\bf C}_{phys}\geq 0
\end{equation}
\begin{equation} \label{lin4'}
{\bf e}^T_N{\bf C}_{phys}\leq n
\end{equation}

It turns out, as in Section~\ref{optim}, that this task can also be (approximately) converted to the maximization 
of a concave function over a convex domain. This convex setting yields that in the optimal solution the ${\bf C}_{phys}$ vector will be 
at a vertex of the polyhedral feasible domain. This forces it to be a 0-1 vector, thus identifying 
(via its 1-coordinates) the optimal choice of the $n$ physical entities, out of the potential $N$, i.e., the  optimal configuration 
of the reconfigurable physical network.

%%%%%%%%%%%%%%%%%%%%%%%%%%%%%%%%%%%%%%%%%%%%%%%%%%%%%%%%%%%%%%%%%%%%%%%

\section{Conclusion}

We have introduced a general and comprehensive model for the design and optimization of Network Slicing and Virtual Networks.
The model is flexible, so that by adjusting some of its elements, it can accommodate  many different specific cases
of importance. Yet, surprisingly, it still allows efficient optimization, in the sense that the global optimum can be well approximated
by efficient algorithms. In the present paper our goal was to describe the conceptual model, elaborate its fundamental equations, as well as issues 
related to its efficient optimization. Numerical investigations and simulation will be the subject of future papers.

\noindent
\appendix[Proofs of Theorems 1 and 2 (Sketch)]

\medskip

 The modification we use in the objective function lies in adding a term that is characteristic to
{\em  link  utilization}.
%  and   makes  the  objective  function
%asymptotically concave. 
Adopting ideas from   Kelly  \cite{kelly2} and  from our previous work Farag\'o et al. \cite{jsacneur, jsac}
(where we used it in a more restricted context),  let  us  define  the {\em utilization function}  $U(y,C_j)$ on a logical entity $j$ of capacity 
$C_j$ by  the {\em implicite}  relation
$$ U(-\log (1-F_j(\rho_j,C_j)),C_j)=\rho_j (1-F(\rho_j,C_j)).$$
$U(y,C_j)$ is  exactly the  mean amount of capacity  in use,  i.e.\ the
average  occupancy,  when  the  loss   probability is $1-\exp
(-y).$ In other words, $U(y,C_j)$ is the average utilization of the logical entity
with respect to a logarithmically scaled loss probability.

Consider a logical entity of capacity $C_j$ and blocking probability $B_j.$ Let
us define the
{\em utilization measure} as
$$ \tilde{U}(B_j,C_j)=\int_0^{-\log (1-B_j)} U(z,C_j)\:dz.$$
Note that the value $y=-\log (1-B_j)$ is a logarithmic measure of 
loss. If $B_j=0$ then $y=0,$ if $B_j$ approaches 1 then $y$ tends to
infinity and the mapping between $B_j$ and $y$ is strictly increasing.

Now we state our modified optimization problem as follows.
\begin{equation} \label{modi}
{\rm Maximize}\;\;\; \sum_r \nu_r \prod_j (1-B_j)^{A_{jr}}+
\sum_j \tilde{U}(B_j,C_j)
\end{equation}
\hspace*{6mm}Subject to
$$ %\begin{equation} %\label{rhoi}
\rho_i = (1-F_i(\rho_i, C_i))^{-1} 
\sum_{r} A_{ir} \nu_r\prod_j \Big( 1-F_j(\rho_j,C_j)\Big)^{A_{jr}}, \;\;\;\;\; i=1,\ldots,m,
$$ %\end{equation}
$$ %\begin{equation} %\label{lin1}
\;\;\;\; {\bf S}{\bf C}\leq {\bf C}_{phys}
$$ %\end{equation}
$$ %\begin{equation} %\label{lin2}
{\bf C}\geq 0
$$ %\end{equation}

%%%%%%%%%
%$${\rm Subject} \;{\rm to}\;\;\;S C\leq C_{phys}\;\,{\rm and}\;\,C
%\geq 0,$$
%where the dependence of $B_j$ on $C_j$ is defined by
%the Erlang fixed point equations
%\begin{equation} \label{erfix}
% B_j=E\left((1-B_j)^{-1} \sum_r A_{jr} \nu_r \prod_i
%(1-B_i)^{A_{ir}},\; C_j\right).
%\end{equation}

{\em Remark:}
The difference between the original and the modified problem is the
additional utilization term in the objective function. Note,
however, that this difference is quite negligible for realistic
values. This follows from the following argument.

Let $T$ be the total carried traffic. Further, let $T'$ be a  weighted
version of the total carried traffic such that the carried traffic  on
each logical route is multiplied by the  length of the route (number of  logical entities
on the route). For  example, if a route  r carries 5 units  of traffic
and r consists  of 3 logical entities,  then r contributes  5 to $  T$ and 15  to
$T'$.

Now we can observe that $T'$  equals to the sum of logical entity  utilizations,
summed over all logical entities, since the  total number of circuits in use  (on
the average) is  exactly the average  carried traffic if  we take into
account that a route occupies capacity on a number of logical entities, equal  to the length of the route.

Let $L$ be the length of the longest route. Then $T'\leq LT.$ Now,  if
we approximate $-\log (1-B)$ by $B,$ which is good for small values of
$B,$ and use the fact that $U(y,C)$ is increasing in $y,$ then we  can
bound the link utilization term  in the objective function from  above
by $BLT,$ where $B$ is the largest blocking probability. In other
words,  instead  of  optimizing  the  total  carried  traffic  $T,$ we
optimize a quantity $Q$ that satisfies
                          $$ T < Q < (1+BL)T.$$
If $B$ is small and $L$ is not too large, then $1+BL$ can be
quite close to 1. For example, if $B < 0.5\%$ and $L=4,$ then we
have
                          $$ T < Q < 1.02 T.$$
Thus, we are optimizing an objective  function that is a
good approximation of the total carried traffic. Extending this informal argument to the general case, we obtain a proof of (ii).

To prove (i), 
 let us  define an auxiliary function $\phi(C)$ as follows.
\begin{equation} \label{fi}
\phi(C)=\min_{y\geq 0}\left\{\sum_r \nu_r
e^{-\sum_jy_jA_{jr}}+\sum_j\int_0^{y_j} U(z,C)\:dz\right\},
\end{equation}
where $y=(y_1,\ldots,y_J).$ 
Now, using the methods of Kelly \cite{kelly2}, we can prove 
the following:

\medskip\noindent
{\em Lemma 1.}\,\,{\em $\phi(C)$ is an asymptotically concave function
of
$C.$}

\medskip\noindent
{\em Lemma 2:}\,\,{\em The logical capacity vector $C$ is an optimal
solution to problem {\rm (\ref{modi})} if and only if it is an optimal
solution to the problem}
\begin{equation} \label{fimax}
{\rm Maximize}\;\;\;\; \phi(C)
\end{equation}
$${\rm Subject} \;{\rm to}\;\;\;S C\leq C_{phys}\;\,{\rm and}\;\,C\geq
0.$$

\medskip  Observe that  by   Lemma 1,  problem   (\ref{fimax})  means  the
maximization  of  an  asymptotically  concave  function  over a convex
domain. Thus, in the asymptotic sense, that is, when the  capacities
are large, the task tends to an ordinary convex programming  task
that  can  be  solved  by  standard  techniques.  The  result yields a
solution to problem (\ref{modi}), as well, by Lemma 2.

Of course, to carry out the optimization, we need to be able to
compute the value of $\phi(C)$ efficiently for any given $C.$
This can be done again by convex optimization, since the computation
of $\phi(C)$  for a given $C$ requires the minimization of a
convex function over a convex domain. This is so, because one can directly check that 
the argument of minimization in (\ref{fi}) is a convex function 
of $y$. 

%For large capacities the link utilization function can be %approximated
%by a  simple formula,  because it  is shown  in \cite{kelly2} 
% that $$
%U(y,C)=C-(e^y-1)^{-1}+o(1) $$ holds as $C$ tends to infinity.

\hfill $\diamondsuit$

\ifCLASSOPTIONcaptionsoff
  \newpage
\fi

% trigger a \newpage just before the given reference
% number - used to balance the columns on the last page
% adjust value as needed - may need to be readjusted if
% the document is modified later
%\IEEEtriggeratref{8}
% The "triggered" command can be changed if desired:
%\IEEEtriggercmd{\enlargethispage{-5in}}

% references section

% can use a bibliography generated by BibTeX as a .bbl file
% BibTeX documentation can be easily obtained at:
% http://mirror.ctan.org/biblio/bibtex/contrib/doc/
% The IEEEtran BibTeX style support page is at:
% http://www.michaelshell.org/tex/ieeetran/bibtex/
%\bibliographystyle{IEEEtran}
% argument is your BibTeX string definitions and bibliography database(s)
%\bibliography{IEEEabrv,../bib/paper}
%
% <OR> manually copy in the resultant .bbl file
% set second argument of \begin to the number of references
% (used to reserve space for the reference number labels box)

\end{document}